\documentclass[12pt]{article}
\usepackage{pdproc,epsfig} 

\begin{document}

\renewcommand{\thefootnote}{\alph{footnote}}
  
\title{
 NEUTRINOS: {\em ``...ANNUS MIRABILIS"}}

\author{A. YU. SMIRNOV}

\address{ International Centre for Theoretical Physics, Strada Costiera 11,\\
31014 Trieste, Italy,  and\\
Institute for Nuclear Research,  Russian Academy of Sciences,Moscow, Russia\\
{\rm E-mail: smirnov@ictp.trieste.it}}

\abstract{Main results and achievements  of 2002 - 2003 
in neutrino physics are 
summarized. The field moves quickly to  new phase with clear 
experimental and  phenomenological programs,  and probably,  
with new theoretical puzzle which may lead us 
to  discoveries of the fundamental importance. 
One of the main results is amazing pattern of
the lepton mixing which emerges from the data. 
The key questions are:  
Does lepton mixing imply new symmetry of Nature?
Is the large (maximal?) mixing related to degeneracy of the neutrino 
mass spectrum?  In this connection priorities of the future studies are 
formulated.}
   
\normalsize\baselineskip=15pt

%%%%%%%%%%%%%%%%%%%%%%%%%%%%%%%%%%%%%%%%%%%%%%%%%%%%%
\section{One year after}
%%%%%%%%%%%%%%%%%%%%%%%%%%%%%%%%%%%%%%%%%%%%%%%%%

\noindent
Year 2002 started by the SNO publication of the direct evidence 
for  the solar neutrino 
flavor conversion\cite{snof} and finished by an announcement  of the first 
KamLAND result\cite{KL} has been  called  
in Ref.\cite{bari} the {\it ``annus mirabilis"} of the 
solar neutrino physics. In 2002 the pioneering works on the detection of solar 
neutrinos have  gotten the highest appreciation  
and in the same year the solar neutrino problem 
(which was the outcome of this detection and the driving force of 
developments in neutrino physics during last 35 years) has been 
essentially resolved. The beginning and the end  have met. 
What happened after? What is an impact of the  {\it annus mirabilis} on 
the field? \\ 

{}From the scientific calendar starting 1 year back from now\footnote{Talk 
given at the 2nd Int. Workshop on Neutrino oscillations in Venice (NOVE) 
December 3-5, 2003, Venice, Italy}:

\begin{itemize}

\item
{\em December 4, 2002.}~  K2K~\cite{K2K}:``Indication of neutrino 
oscillations in a 250 km long baseline experiment''.

\item
{\em December 6, 2002.} KamLAND~\cite{KL}: ``First Results from 
KamLAND: Evidence  for Reactor for reactor anti-neutrino 
disappearance''.

\item
{\it December 10, 2002.}~ The ceremony of the Nobel Prize award:
R. Davis Jr. and M. Koshiba: ``... for pioneering
contribution to astrophysics, in particular for the detection of cosmic
neutrinos''.

\item
{\em February 11, 2003.}~  WMAP~\cite{WMAP}: ``First year Wilkinson 
microwave anisotropy probe
observations: determination of cosmological parameters...''

\item
{\em September 3, 2003.}~SuperKamiokande-I\cite{SK}: ``Precise measurement of the 
solar neutrino Day/Night and seasonal variations in 
Super-Kamiokande-I. 

\item
{\em September 7, 2003.}~ SNO salt phase results~\cite{salt}: 
``Measurements of the total active B-8 solar neutrino flux 
at the Sudbury neutrino observatory with enhanced neutral current sensitivity''.

\item
{\em October 27, 2003.}~ Sloan Digital Sky Survey~\cite{SDSS}: 
``Cosmological parameters from SDSS and WMAP''. 

\end{itemize}

\noindent
There is a number of immediate consequences of these results: 

%\noindent
1). The LMA MSW solution of the solar neutrino problem is confirmed. 

%\noindent
2). The oscillation parameters $\Delta m^2_{12}$ and
$\theta_{12}$ are determined with reasonable accuracy.
In particular, significant deviation 
of the 1-2 mixing from maximal is established. 

%\noindent
3). The  key step is done in the reconstruction of the neutrino mass and 
flavor spectrum. The dominant structures of the mixing matrix and both 
$\Delta m^2$  (apart from the sign of $\Delta m^2_{13}$) are known. 

%\noindent
4). The confirmation of the LMA solution opens 
a possibility to measure the CP violation in
the leptonic sector in the future LBL experiments.

%\noindent
5). In the connection to  LMA,   a possibility of 
substantial cancellation of contributions in the neutrinoless double beta
decay is confirmed. This, in turn, has serious impact 
on perspectives of determination  of the absolute scale of 
neutrino mass and the role of the Majorana phases. 

%\noindent
6).  Picture of the flavor conversion of neutrinos from SN 1987A 
is determined.  

%\noindent
7).  The LMA oscillations of the atmospheric neutrinos should exist. 
This opens new possibility to search for the deviation of 2-3 mixing 
from maximal. 

%\noindent
8).  Strong bound on the leptonic asymmetry of the Universe is 
established. 

%\noindent
9).  The first KamLAND result marks the birth of neutrino geophysics. 

%\noindent
10).  Important  cosmological bound on neutrino mass is given. \\

%\noindent
%11).  New theoretical puzzle  emerges?  

These results  moved the field to new phase with new goals, 
experimental programs and theoretical problems. 

%%%%%%%%%%%%%%%%%%%%%%%%%%%%%%%%%%%%%%%%%%%%%%%%%%%%%%%%%%%%%%%
\section{Summarizing achievements}
%%%%%%%%%%%%%%%%%%%%%%%%%%%%%%%%%%%%%%%%%%%%%%%%%%%%%%%%%%%%%%%

%%%%%%%%%%%%%%%%%%%%%%%%%%%%%%%%%%%%%%%%%%%%%%%%%%%%
\subsection{After SNO salt results}
%%%%%%%%%%%%%%%%%%%%%%%%%%%%%%%%%%%%%%%%%%%%%%%%%%%%

\begin{figure}[h!]
\begin{center}
\hspace{-1cm}\epsfxsize8.5cm\epsffile{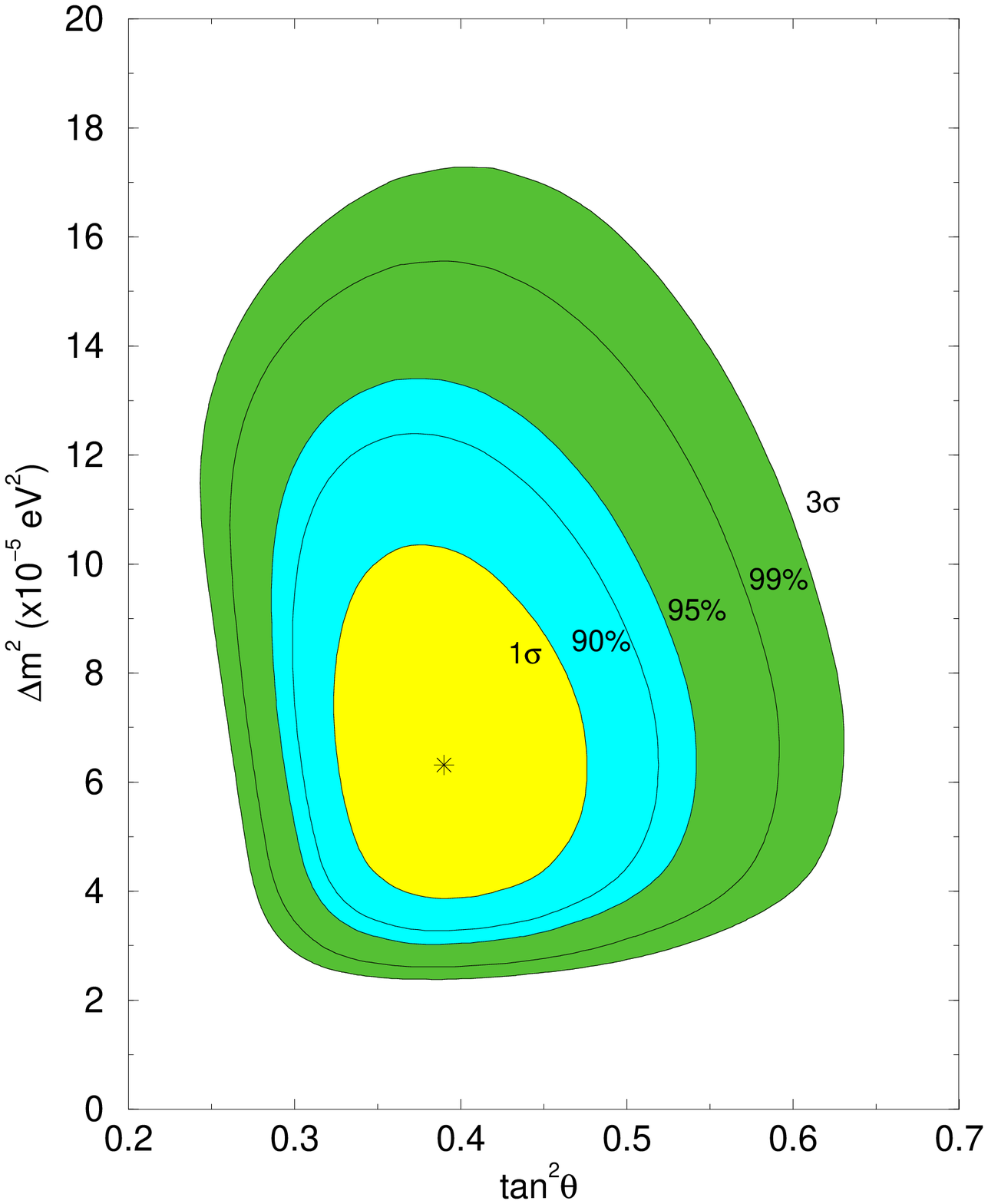} \hskip -1cm 
\epsfxsize8.5cm\epsffile{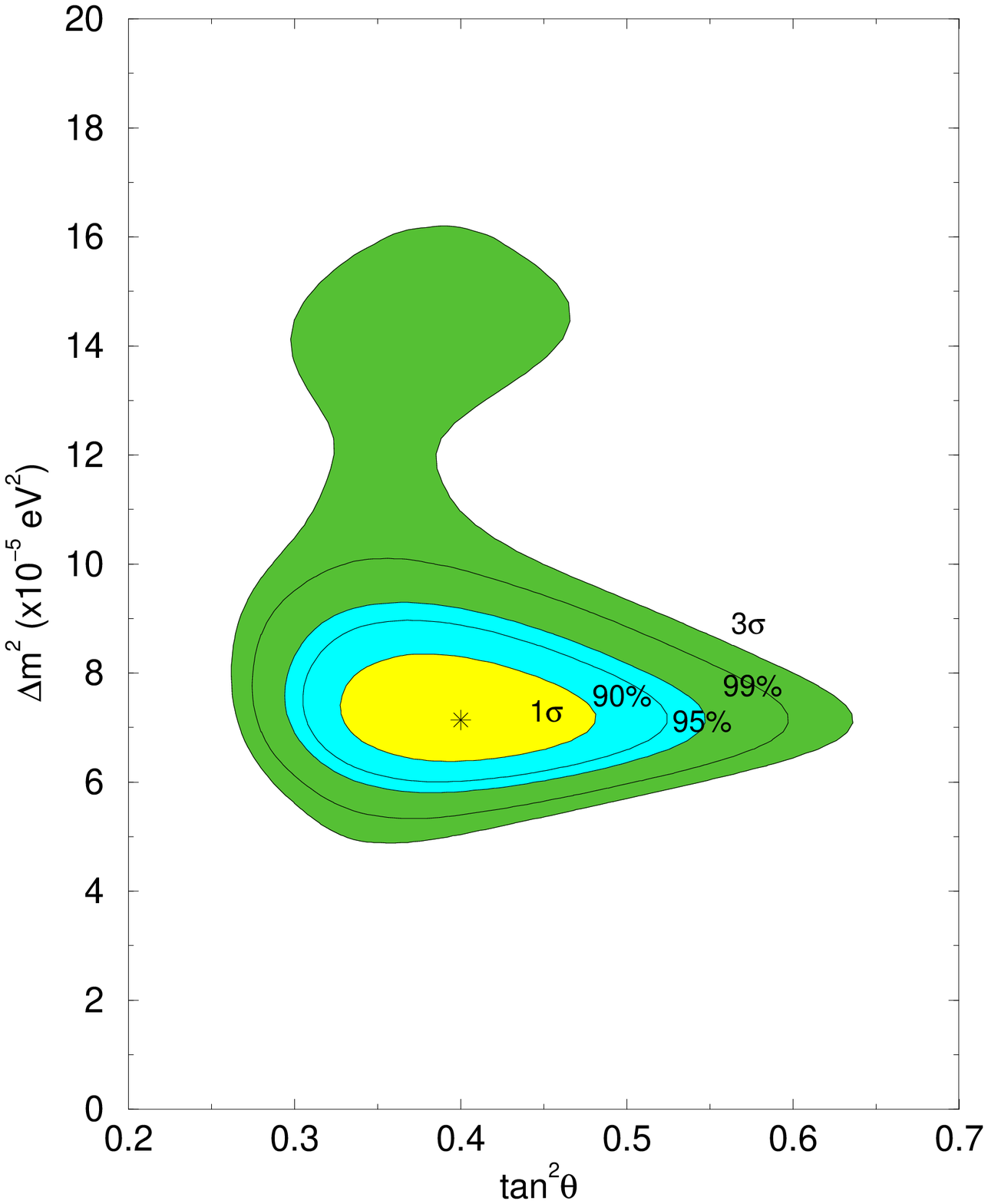}
\vspace{-1.5cm}
\caption{The allowed regions of oscillation parameters 
from the 2$\nu$ analysis of the solar neutrino data (left) and 
the combined fit of the solar 
neutrino data and the KamLAND spectrum (right) 
at $1\sigma$, $2\sigma$, $3\sigma$ CL.$^{15)}$}
\label{allowed}
\end{center}
\end{figure}

The SNO  salt phase results\cite{salt}
have  further confirmed the correctness of the Standard Solar Model (SSM) neutrino 
fluxes\cite{ssm} and the realization of the MSW large mixing (LMA) conversion 
mechanism\cite{msw}  inside the Sun.\cite{balan}$^{-}$\cite{pedro3} 
%~\cite{salt,balan,fogli,valle2,alia,crem,choubey,pedro3}. 
In Fig:~\ref{allowed} we show the allowed region of  the oscillation 
parameters  $\tan^2 \theta_{12}$ and $\Delta m^2_{12}$ from 
the  $2\nu$ analysis of the solar neutrino data (left)
and from the combined analysis of the solar neutrino and KamLAND\cite{KL} results 
(right). 
The best-fit values of the parameters are    
\begin{equation}
\Delta m^2_{12} = 7.1 \times 10^{-5} {\rm eV}^2, ~~ 
\tan^2 \theta_{12} = 0.4. 
\label{eq:bf}
\end{equation}

%%%%%%%%%%%%%%%%%%%%%%%%%%%%%%%%%%ffff1%%%%%%%%%%%%%%%%%%%%%%%%%%%%%%%%%%%%%
%%%%%%%%%%%%%%%%%%%%%%%%%%%%%%%%%%%%%%%%%%%%%%%%%%%%%%%%%%%%%%%%%%%%%%%
%\vspace{-5mm}
%%
Combined fit of the solar, KamLAND\cite{KL} and CHOOZ\cite{chooz} results favors nearly  
zero 1-3 mixing: $\sin^2 \theta_{13} \sim 0$\rlap{.}\,\cite{choubey,pedro3} 
Implications of these results can be formulated in the following way.

1). The l-LMA   region with $\Delta m^2_{12} < 10^{-4}$ ${\rm eV}^2$ is selected,  and 
the h-LMA region is accepted now at $3\sigma$ only.    

2). Maximal 1-2 mixing is strongly disfavored. The upper bound is   
\begin{equation}
\tan^2 \theta_{12} <  0.64 ~~~~~(3 \sigma). 
\label{eq:up12}
\end{equation}
That is, significant deviation of the 1 - 2 mixing from maximal is established which can be 
expressed as 
\begin{equation}
1/2 - \sin^2\theta_{12} \sim \sin^2\theta_{12}. 
\end{equation}

3). As a result of more precise determination of the oscillation parameters the 
physics of the conversion is  now  determined 
quantitatively\rlap{.}\,\cite{fogli,pedro3} 
In particular, recent results  show  relevance of the notion of resonance,  they 
fix   the relative strength of the effects of the adiabatic conversion 
and the oscillations as function of the neutrino energy~\cite{pedro3}.  

Next KamLAND data release  is extremely important for understanding stability of the results, 
backgrounds, contribution of the geo neutrinos and more precise determination of parameters.   
  
Concerning  potential problems of the LMA solution, namely,  the low Homestake 
rate~\cite{Cl} and the absence of the upturn of the boron neutrino spectrum at 
low energies:   
Recent measurements of the nuclear cross-sections by LUNA
experiment lead to decrease of the CNO fluxes~\cite{luna}, and consequently,  reduced
difference of the Homestake result and the LMA prediction~\cite{ster}.  
Forthcoming SNO spectral results may shed some light on  existence of the 
upturn~\cite{ster}.

%%%%%%%%%%%%%%%%%%%%%%%%%%%%%%%%%%%%%%%%%%%%%%%%%%%%%%%%%%%%%%%
\subsection{Atmospheric neutrinos and 2-3 mixing}
%%%%%%%%%%%%%%%%%%%%%%%%%%%%%%%%%%%%%%%%%%%%%%%%%%%%%%%%%%%%%%%%%

A recent refined analysis of the SuperKamiokande data in terms 
of $\nu_{\mu} - \nu_{\tau}$ oscillations gives\cite{atm} at 90 \% C.L. 
\begin{equation}
\Delta m^2_{13} = (1.3 - 3.0) \times 10^{-3} {\rm eV}^2, ~~
\sin^2 2\theta_{23} >  0.91 ~
\label{eq:atm}
\end{equation}
with the best fit at $\Delta m^2_{12} = 2.0 \times 10^{-3}$ eV$^2$ and $\sin^2 2\theta_{12} = 1.0$. 
Combined analysis of the CHOOZ and the atmospheric neutrino data puts the upper bound  
on the  1-3 mixing\cite{atmfo} 
\begin{equation}
\sin^2 \theta_{13} <   0.067 ~~~~(3\sigma). 
\label{eq:atm}
\end{equation}

\begin{figure}[h!]
\begin{center}
\hspace{-1cm} \epsfxsize14cm\epsffile{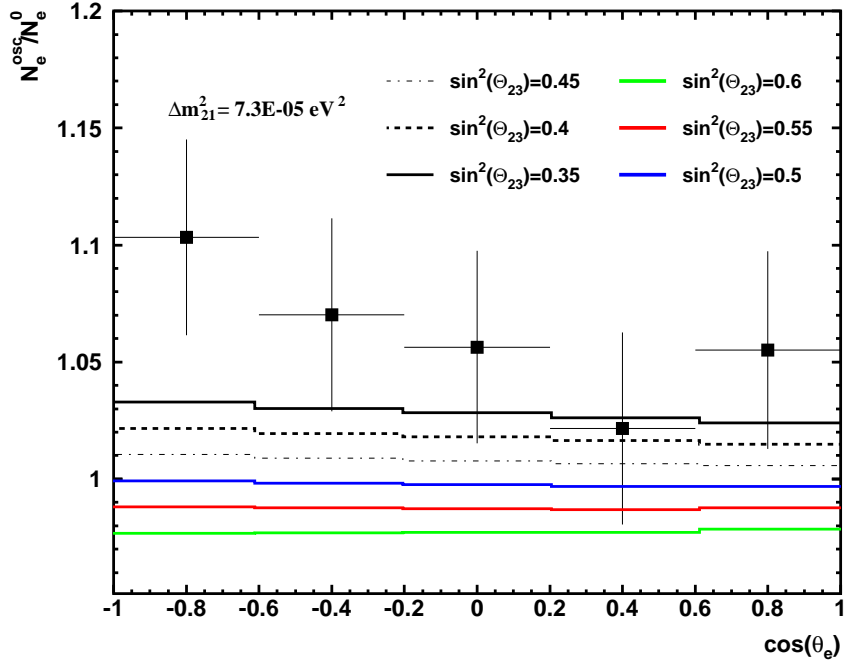}
\vskip -10cm
\caption{The ratio of numbers of the $e$-like events with and without oscillations 
as function of the zenith angle of the electron for  
different values of $\sin^2 \theta_{23}$.$^{22)}$ 
Other parameters are $\sin^2 2\theta_{12} = 0.82$, $\sin \theta_{13} = 0$ and 
$\Delta m^2_{12} = 7.3 \times 10^{-5}$ eV$^2$.  Also  shown are the 
SuperKamiokande experimental points. }
\label{atm}
\end{center}
\end{figure}

The open question is whether oscillations of the 
atmospheric $\nu_e$ exist? There are two possible sources of these oscillations: 
(i) non-zero 1-3 mixing and ``atmospheric" $\Delta m^2_{13}$, and 
(ii) solar oscillation parameters in Eq.~(\ref{eq:bf}).  
Also their interference  should exist\rlap{.}\,\cite{PS-L} After  confirmation of the 
LMA-MSW solution we can definitely say that oscillations driven by the LMA parameters 
(the LMA oscillations) should show up at some level. 
Relative change of the $\nu_e$ flux due to the LMA oscillations can be 
written as\cite{PS-L} 
\begin{equation}
\frac{F_e}{F_e^0} - 1 = P_2 (r \cos^2 \theta_{23} - 1), 
\label{eq:atm-e}
\end{equation}
where $P_2(\Delta m^2_{12}, \theta_{12})$ is the $2\nu$ transition probability and 
$r \equiv {F_{\mu}^0}/{F_e^0}$ is the ratio of the original $\nu_{\mu}$ and $\nu_e$ fluxes. 
In the sub-GeV region, where $P_2$ can be of the order 1, the ratio equals $r \approx 2$,  so that 
the oscillation effect is proportional to the deviation of the 2-3 mixing from the
maximal value: 
\begin{equation}
D_{23} \equiv 1/2 - \sin^2 \theta_{23}
\label{devia}
\end{equation}
In  Fig:~\ref{atm}  from \cite{PS-L} 
we show the ratio of numbers of the $e$-like events with and without oscillations
as function of the zenith angle of the electron. 
For the allowed range  of $\sin^2 \theta_{23}$ and the present best-fit value of 
$\Delta m^2_{12}$ the excess can be as large as 5 - 6\%. 
The excess increases with decreasing energy.  
\vskip 0.7cm
%%%%%%%%%%%%%%%%%%%%%%%%%%%%%%%%%%ffff2%%%%%%%%%%%%%%%%%%%%%%%%%%%%%%%%%%%%%
%%%%%%%%%%%%%%%%%%%%%%%%%%%%%%%%%%%%%%%%%%%%%%%%%%%%%%%%%%%%%%%%%%%%%%%

\vskip -0.5cm

Future searches for the excess can be  used to restrict or measure $D_{23}$. 
In fact, the latest analysis, 
(without  renormalization of the original fluxes) shows some excess of the $e$-like events at 
low energies and the absence of excess in the multi-GeV sample, thus  giving a
hint of non-zero $D_{23}$.  
Establishing this deviation has important consequences for understanding the origins 
of neutrino masses and mixing. 
%%
%%%%%%%%%%%%%%%%%%%%%%%%%%%%%%%%%%ffff3%%%%%%%%%%%%%%%%%%%%%%%%%%%%%%%%%%%%%
\begin{figure}[h!]
\begin{center}
\hspace{-0.1cm} \epsfxsize11cm\epsffile{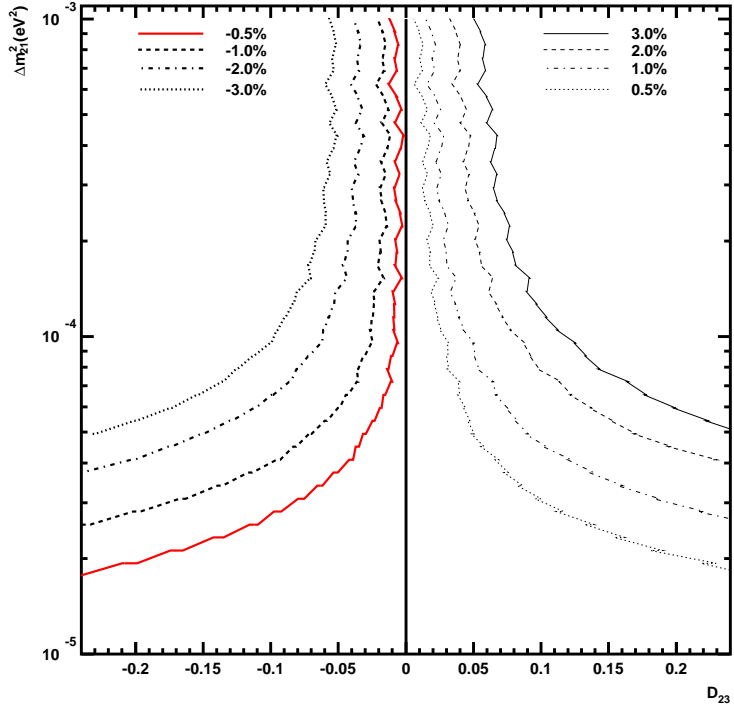}
\vskip -3cm
\caption{Contours of constant excess of the e - like events in the 
$\Delta m^2_{12}$-$D_{23}$ plane.$^{22)}$}
\label{cont}
\end{center}
\end{figure}
%%%%%%%%%%%%%%%%%%%%%%%%%%%%%%%%%%%%%%%%%%%%%%%%%%%%%%%%%%%%%%%%%%%%%%%

In Fig:~\ref{cont}  we  show the contours of constant excess of the e - like 
events for zero  1-3 mixing. According to this figure, establishing the excess 
at the level
$3\%$ would imply for  $\Delta m^2_{12} = 7\cdot 10^{-5}$ eV$^2$ the lower bound 
on the deviation: $D_{23} > 0.17$. 
The 1-3 mixing generates the interference effect  
between the LMA oscillations amplitudes\rlap{.}\,\cite{PS-L}  
The interference contribution  does not contain the ``screening" factor, 
as in Eq.~(\ref{eq:atm-e}), and can reach 2\,--\,4\% 
for the allowed values of $\sin \theta_{13}$.  
This produces an uncertainty in the determination 
of $D_{23}$ which can be reduced once  stronger 
bound on 1-3 mixing is obtained.  
In any case observation of the excess of e- like events  
at the level of $\sim 5\%$ will  imply
strong  deviation of the 2-3 mixing from the maximal one.

%%%%%%%%%%%%%%%%%%%%%%%%%%%%%%%%%%%%%%%%%%%%%%%%%%%%%%%%%%%%%%%%%%%%%%%%%%%%%%%%%%%%%%%
\subsection{Neutrinos from SN1987A: flavor conversion}
\label{subsec:sol}
%%%%%%%%%%%%%%%%%%%%%%%%%%%%%%%%%%%%%%%%%%%%%%%%%%%%%%%%%%%%%%%%%%%%%%%%%%%%%%%%%%%%%%%

After confirmation of the LMA-MSW solution we can definitely say that the
effect of  flavor conversion has already been observed  in 1987. 
One must take into account the conversion effects in analysis of 
SN1987A\cite{sn87a} and future  supernova  neutrino data. 

%%%%%%%%%%%%%%%%%%%%%%%%%%%%%%%%%%ffff4%%%%%%%%%%%%%%%%%%%%%%%%%%%%%%%%%%%%%
\begin{figure}[h!]
\begin{center}
\hspace{-0.1cm} \epsfxsize12cm\epsffile{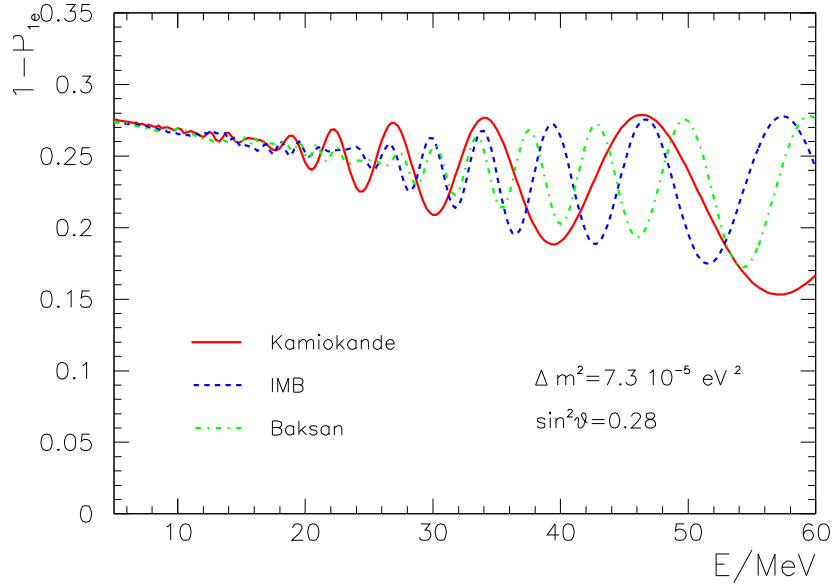}
\vskip -0.5cm
\caption{The permutation factor $\bar p = 1 - P_{1e}$ 
as a function of neutrino energy for 
Kamiokande II, IMB and Baksan detectors.$^{26)}$ }
\label{sn87}
\end{center}
\end{figure}
%%%%%%%%%%%%%%%%%%%%%%%%%%%%%%%%%%%%%%%%%%%%%%%%%%%%%%%%%%%%%%%%%%%%%%%
%%
In terms of the original fluxes of the electron,  
$F^0(\bar \nu_e)$, and muon, $F^0(\bar \nu_{\mu})$,   antineutrinos,  
the electron antineutrino flux at the detector can be written as 
\begin{equation}
F(\bar \nu_e) = F^0(\bar \nu_e) + \bar{p} \Delta F^0,  
\label{flu}
\end{equation}
where $\Delta F^0 \equiv F(\bar \nu_{\mu}) -  F(\bar \nu_e)$, 
and $\bar{p}$ is the permutation factor. In  assumptions of the  normal mass hierarchy 
(ordering) and the absence of new neutrino states, $\bar{p}$ can be calculated 
precisely: $\bar{p} = 1 - P_{1e}$,  where $P_{1e}$ is  
the probability of $\bar{\nu}_1 \rightarrow \bar{\nu}_e$ transition 
inside the Earth\rlap{.}\,\cite{DS,LS-87} 
It can be written as $\bar{p} = \sin^2 \theta_{12} - f_{reg}$, where $f_{reg}$ describes  the effect of 
oscillations (regeneration of the $\bar{\nu}_e$ flux) inside the Earth. 
Due to the difference in distances traveled by neutrinos to 
Kamiokande, IMB and Baksan detectors inside the Earth:   
4363~km, 8535~km and 10449~km correspondingly,  the permutation factors differ   
for these detectors (Fig:~\ref{sn87}).  
The Earth matter effect can partially explain the difference between the 
Kamiokande and the IMB spectra of events\rlap{.}\,\cite{LS-87} 

%%%%%%%%%%%%%%%%%%%%%%%%%%%%%%%%%%ffff5%%%%%%%%%%%%%%%%%%%%%%%%%%%%%%%%%%%%%
\begin{figure}[h!]
\begin{center}
\hspace{-0.1cm} \epsfxsize12cm\epsffile{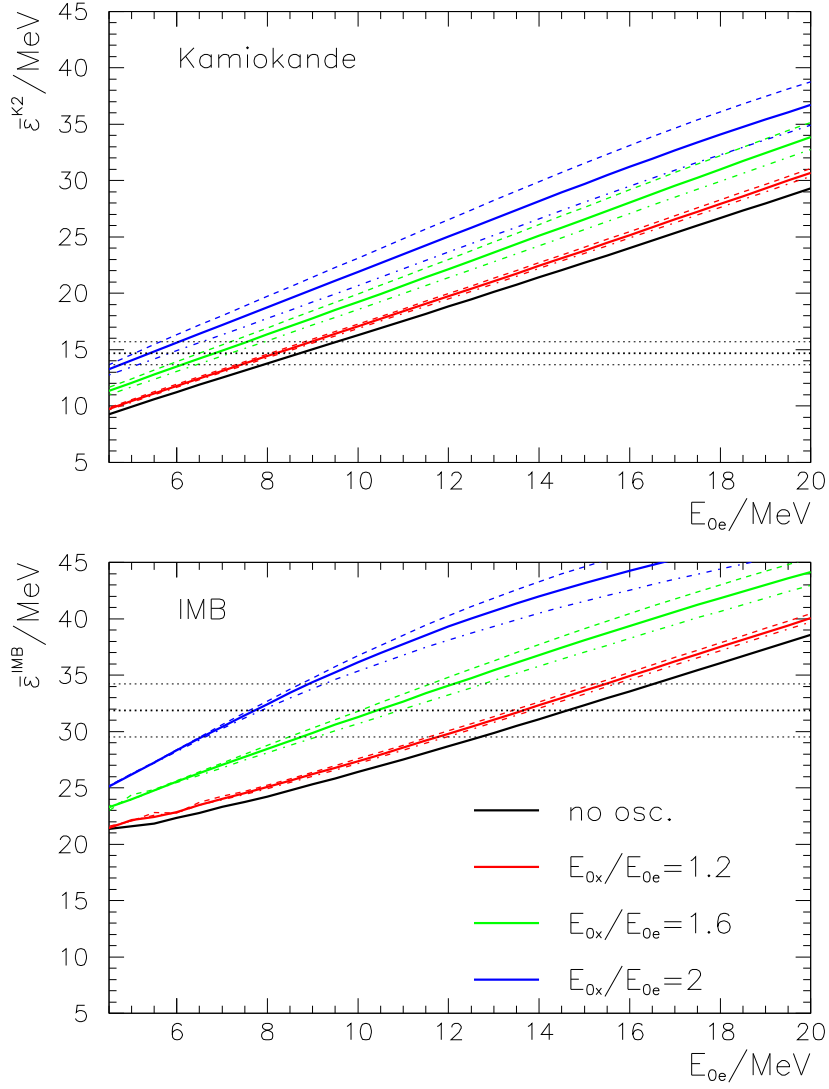}
\vskip -0.5cm
\caption{The dependence of the average energy of the 
observed events on the average energy of the original $\bar{\nu_e}$ spectrum $E_{0e}$ for
different values of $r_E \equiv E_{0\mu}/E_{0e}$ and $r_L \equiv L_{0\mu}/L_{0e}$
for Kamioka-2 (upper panel) and IMB (lower panel)$^{22)}$.}
\label{energy}
\end{center}
\end{figure}
%%%%%%%%%%%%%%%%%%%%%%%%%%%%%%%%%%%%%%%%%%%%%%%%%%%%%%%%%%%%%%%%%%%%%%%
%%
In contrast to $\bar p$, the effect of conversion on
the observed signals depends substantially on the
parameters of  original neutrino spectra. As an illustration,
in Fig:~\ref{energy} we show  the average energy of the observed
events as a function  the average energy of the original $\bar{\nu}_e$ spectrum 
$E_{0e}$ for
different values of $r_E \equiv E_{0\mu}/E_{0e}$ and $r_L \equiv L_{0\mu}/L_{0e}$
for Kamioka-2 (upper panel) and IMB (lower panel)~\cite{LS-87a}.
Notice that conversion can lead to   increase of the average energy by 
(30 - 40)\%.  Inversely, as follows from Fig:~\ref{energy},  not taking into account 
the conversion can lead to  errors in determination of the average energy of 
original spectrum  of the order  40 - 50 \% in K2, and  factor of 2 in IMB. 

For the inverted mass hierarchy and  $\sin^2 \theta_{13} > 10^{-4}$ 
one would get nearly complete  permutation,    
$\bar{p} \approx 1$, and therefore a harder $\bar\nu_e$ spectrum,
as well as  the  absence of the Earth matter effect. This is disfavored by  the 
data\rlap{,}\,\cite{sn-inv}  
though in view of small statistics and uncertainties in the original fluxes  
it is not possible to make a firm statement. 

%%%%%%%%%%%%%%%%%%%%%%%%%%%%%%%%%%%%%%%%%%%%%%%%%%%%%%%%%%%%%%%%%%%%%%%%%%%%%%%%%%%%%%%
\subsection{Absolute Scale of Mass}
\label{subsec:abs}
%%%%%%%%%%%%%%%%%%%%%%%%%%%%%%%%%%%%%%%%%%%%%%%%%%%%%%%%%%%%%%%%%%%%%%%%%%%%%%%%%%%%%%%

\begin{figure}[h!]
\begin{center}
\hspace{-0.1cm} \epsfxsize9cm\epsffile{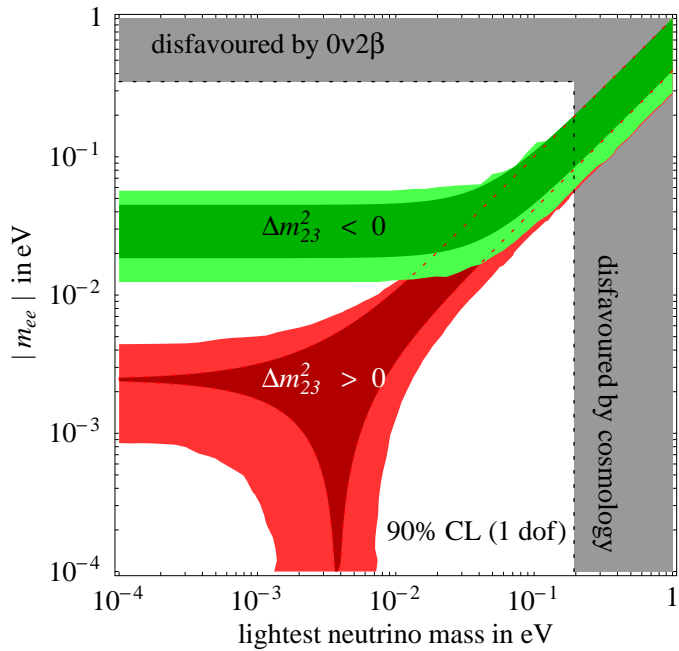}
\caption{The 90\% CL range for $m_{ee}$ as a function of the lightest neutrino mass 
for the normal ($\Delta m_{23}^2 > 0$) and inverted ($\Delta m_{23}^2 < 0$) mass 
hierarchies $^{28)}$.
The darker regions show how the allowed 
range for the present best-fit values of the parameters 
with negligible errors.    
}
\label{bb}
\end{center}
\end{figure}

{}From the oscillation results  we can put a lower limit on the heaviest 
neutrino mass: 
\begin{equation}
m_h  \geq \sqrt{\Delta m^2_{13}} > 0.04~ {\rm eV}, 
\label{eq:ab1}
\end{equation}
where $m_h = m_3$ for the normal mass hierarchy,  and $m_h = m_1 \approx m_2$ for the 
inverted hierarchy. 
The neutrinoless double beta decay is determined by the effective mass  
\begin{equation}
m_{ee} = |\sum_k U_{ek}^2 m_k e^{i\phi(k)}|, 
\label{eq:it}
\end{equation}
where $\phi(k)$ is the Majorana phase of the $k$ eigenvalue. 
Fig:~\ref{bb} from~\cite{vissani} summarizes 
the present knowledge of the absolute mass scale.  
Shown are  the allowed  (at 90 \% CL) regions in the plane of  $m_{ee}$ probed by 
the $\beta\beta_{0\nu}$ decay 
and  the mass of lightest neutrino  probed by the direct kinematical methods and cosmology. 
The best present bound on $m_{ee}$ is given by the Heidelberg-Moscow experiment: 
$m_{ee} < 0.35 - 0.50$ eV\rlap{,}\,\cite{HM-neg} part 
of collaboration claims an evidence of a  positive signal\rlap{.}\,\cite{HM-pos}   
%%%%%%%%%%%%%%%%%%%%%%%%%%%%%%%%%%ffff7%%%%%%%%%%%%%%%%%%%%%%%%%%%%%%%%%%%%%
%%%%%%%%%%%%%%%%%%%%%%%%%%%%%%%%%%%%%%%%%%%%%%%%%%%%%%%%%%%%%%%%%%%%%%%

The present  double beta decay seraches and 
cosmology have similar sensitivities:  $m_{ee} \sim m_1 \sim (0.2 - 0.5)$ eV.
This value corresponds to the degenerate mass spectrum:    
$m_{1} \approx m_2 \approx m_3 \equiv m_0$. 
Analyses of cosmological data (with WMAP)  result in  the 95\% C.L. 
upper bounds $m_0 < 0.23$ eV\rlap{,}\,\cite{cosm}  
$m_0 < 0.6$ eV\cite{cosm1} and $m_0 < 0.34$ eV\rlap{.}\,\cite{cosm2} 
Independent analysis which includes the X-ray galaxy cluster data  gives 
non-zero  value $m_0 = 0.20 \pm 0.10$ eV\rlap{.}\,\cite{cosm3}

%The present upper limit from the direct measurements (tritium decay) will be improved 
%by future KATRIN experiment down to $m \sim 0.25$ eV. 

Future improvements of the upper bound on $m_{ee}$ have the potential to distinguish between the 
hierarchies: 
if the bound $m_{ee} < 0.01$ eV is established, 
the inverted hierarchy will be excluded at 90 \% C.L. (see Fig:~\ref{bb}). 
%Similar bound from:\,\cite{mura} $m_{ee} < 0.015$ eV. 

%%%%%%%%%%%%%%%%%%%%%%%%%%%%%%%%%%%%%%%%%%%%%%%%%%%%%%%%%%%%%%%%%%%%%%
\subsection{Mass Spectrum and Mixing}
%%%%%%%%%%%%%%%%%%%%%%%%%%%%%%%%%%%%%%%%%%%%%%%%%%%%%%%%%%%%%%%%%%%%%%%%

\begin{figure}[h!]
\begin{center}
\hspace{-0.1cm} \epsfxsize12cm\epsffile{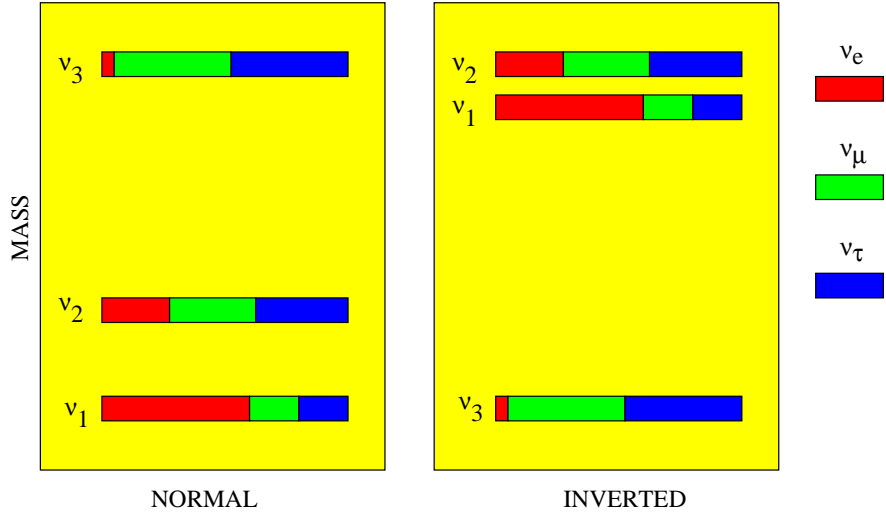}
\caption{Neutrino mass and flavor spectra for the normal (left) and inverted (right) 
mass hierarchies. The distribution of flavors (colored parts of boxes) in the mass eigenstates 
corresponds to the best-fit values of mixing parameters  and  $\sin^2 \theta_{13} = 0.05$. 
}
\label{sp}
\end{center}
\end{figure}

Information obtained 
from the oscillation experiments 
allows us  to make significant progress in the reconstruction of the neutrino mass and flavor  
spectrum  (Fig:~\ref{sp}). 
%%%%%%%%%%%%%%%%%%%%%%%%%%%%%%%%%%ffff4%%%%%%%%%%%%%%%%%%%%%%%%%%%%%%%%%%%%%
%%%%%%%%%%%%%%%%%%%%%%%%%%%%%%%%%%%%%%%%%%%%%%%%%%%%%%%%%%%%%%%%%%%%%%%

Using a global fit of the oscillation data one can find~\cite{concha}  the  (90\% 
CL) intervals for 
the  elements of the PMNS mixing matrix $||U_{\alpha i}||$: 
\begin{equation}
\left(
\begin{tabular}{lll}
0.79 - 0.86 & 0.50 - 0.61 & 0.0 - 0.16\\
0.24 - 0.52 & 0.44 - 0.69 & 0.63 - 0.79\\
0.26 - 0.52  & 0.47 - 0.71 & 0.60 - 0.77\\
\end{tabular}
\right), 
\label{eq:it}
\end{equation}
where columns  correspond to the flavor 
index and rows to the mass index\rlap{.} 

\begin{figure}[h!]
\begin{center}
\hspace{-0.1cm} \epsfxsize9cm\epsffile{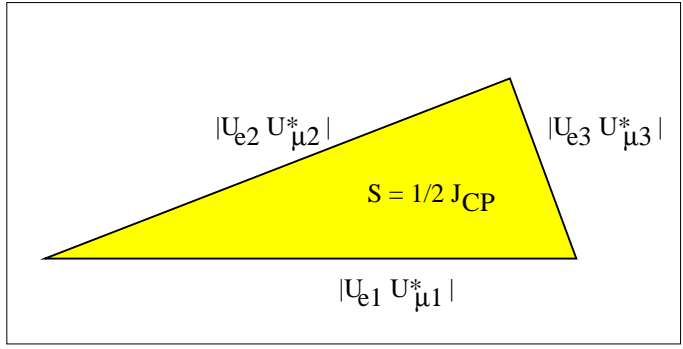}
\caption{Possible leptonic unitarity triangle. We take the best-fit values of  
$\theta_{12}$,  and $\theta_{23}$ and $\sin \theta_{13} = 0.16$.}
\label{tri}
\end{center}
\end{figure}

Now we are in a position to construct the leptonic unitarity triangle, though 
the finite size of one angle and therefore the length of  one is  still 
unknown. For practical 
reason (no intensive $\nu_{\tau}$ beams) we consider the triangle which employs the 
$e$- and $\mu$- rows of the mixing matrix  (Fig:~\ref{tri}).  
The triangle is not degenerate in spite of the strong bound on the 1-3 mixing. 

The area of the triangle is related to  the Jarlskog invariant  
$J_{CP} \equiv Im[{U_{e1} U_{\mu2} U_{e2}^* U_{\mu1}^*}]$:
$S = J_{CP}/2$. Reconstruction of the triangle is complementary 
to  measurements of the neutrino-antineutrino asymmetries in oscillations. 
The main problem here is the coherence. For the triangle method we need to 
study interactions of the mass eigenstates, whereas in practice we deal  with
flavor (coherent) states. So, breaking of the coherence, 
averaging of oscillations, experiments with the beams of mass eigenstates and measurements of the 
survival (rather than transition) probabilities are the key elements of the  
method\rlap{.}\,\cite{yasaman}  

%%%%%%%%%%%%%%%%%%%%%%%%%%%%%%%%%%ffff5%%%%%%%%%%%%%%%%%%%%%%%%%%%%%%%%%%%%%
%%%%%%%%%%%%%%%%%%%%%%%%%%%%%%%%%%%%%%%%%%%%%%%%%%%%%%%%%%%%%%%%%%%%%%%

%%%%%%%%%%%%%%%%%%%%%%%%%%%%%%%%%%%%%%%%%%%%%%%%%%%%%%%%%%%%%%%
\subsection{Main features}
%%%%%%%%%%%%%%%%%%%%%%%%%%%%%%%%%%%%%%%%%%%%%%%%%%%%%%%%%%%%%%%%

Information on neutrino masses and mixing   
can be summarized in the following way.

1). Taking the lower bound  from Eq.(\ref{eq:ab1}) and  the upper cosmological 
bound we can conclude that the absolute scale of neutrino mass
is known within one order of magnitude: 
\begin{equation}
m_h \sim 0.04 - 0.4 ~{\rm eV}.
\end{equation}
This interval is still  too large from the theoretical point of view. 
Depending on specific value of the mass within this interval,  one arrives at  
completely different conclusions.

2). The observed ratio  of the mass squared differences, 
$\Delta m^2_{12}/\Delta m^2_{23} = 0.01 - 0.15$,  
implies that there is no strong hierarchy of neutrino masses: 
\begin{equation}
\frac{m_2}{m_3} > \sqrt{\frac{\Delta m^2_{12}}{\Delta m^2_{23}}} = 
0.18^{+ 0.22}_{-0.08}. 
\label{eq:hie1}
\end{equation}
For charge leptons the corresponding ratio is 0.06.

3). There is the bi-large or large-maximal mixing between the neighboring families
(1 - 2) and (2 - 3). Still  rather significant deviation of the 2-3 mixing 
from the maximal one is possible and it is not excluded, 
{\it e.g.}, that 1-2 and 2-3 are equal. 
Mixing between remote (1-3) families is weak.  

There is interesting and rather precise relation
\begin{equation}
\theta_{12} + \theta_c = \theta_{23} \sim 45^0, 
\end{equation}
where $\theta_c$ is the Cabibbo angle. 
It is not clear if this equality is just accidental coincidence.
It seems  there is no simple scenario which leads to the equality 
$\theta_{12} = 45^0 -  \theta_c$, 
that is, to scenario in which the lepton mixing = bi-maximal mixing 
(which follows from the neutrino mass 
matrix) - corrections (which follow from the charge lepton mass matrix).
Though there is again an unexpected  relation
\begin{equation}
\tan \theta_c \approx \sqrt{\frac{m_{\mu}}{m_{\tau}}} = \lambda \approx 0.2. 
\end{equation}
This may further testify for quark - lepton similarity. 
  
%%%%%%%%%%%%%%%%%%%%%%%%%%%%%%%%%%%%%%%%%%%%%%%%%%%%%%%%%%%%
\subsection{Achieved results: where we are? }
%%%%%%%%%%%%%%%%%%%%%%%%%%%%%%%%%%%%%%%%%%%%%%%%%%%%%%%%%%%%%

The achieved results allow us to formulate clear 
program of further phenomenological and experimental 
studies which includes determination of 

\begin{itemize}

\item
the absolute mass scale $m_1$;

\item
the type of mass spectrum;  
hierarchical; non-hierarchical with certain ordering; degenerate,
which is related to the value of  $m_1$; 

\item
the type of mass  hierarchy (ordering): normal, inverted;  

\item
the 1-3 mixing; 

\item
the CP-violating  Majorana phases; 

\item
deviations of the 2-3 from maximal.    

\end{itemize}

An important issue is searches for new neutrino states.

On the other hand we get new theoretical puzzle which may lead us to new 
fundamental discoveries.  The puzzle is related to unexpected pattern 
of the lepton mixing and possible mass spectrum.

%%%%%%%%%%%%%%%%%%%%%%%%%%%%%%%%%%%%%%%%%%
\section{Toward the underlying physics}
%%%%%%%%%%%%%%%%%%%%%%%%%%%%%%%%%%%%%%%%%%%%%%

The first step in  attempts to uncover the underlying physics 
could be reconstruction of the neutrino mass matrix 
and studies of its properties. Here we assume  
that neutrinos are the Majorana particles. 

%%%%%%%%%%%%%%%%%%%%%%%%%%%%%%%%%%%%%%%%%%%%%%%%%%%%%%
\subsection{ Neutrino mass matrix}
%%%%%%%%%%%%%%%%%%%%%%%%%%%%%%%%%%%%%%%%%%%%%%%%%%%%%

The Majorana mass matrix of neutrinos in the flavor basis can be written as 
\begin{equation}
m = U^* m^{diag} U^+,
\label{eq:mass}
\end{equation}
where $U = U(\theta_{ij}, \delta)$ 
is the mixing matrix, $\delta$ is the Dirac CP-violating phase,   
and 
\begin{equation}
m^{diag} \equiv diag (m_1 e^{-2i\rho},~ m_2,~ m_3 e^{-2i\sigma}).  
\label{eq:mass}
\end{equation}
Here $\rho$ and $\sigma$ are the Majorana phases. 
The mass eigenvalues equal $m_2 = \sqrt{m_1^2  + \Delta m^2_{12}}$, and
$m_3 = \sqrt{m_1^2  + \Delta m^2_{13}}$. 

\begin{figure}[h!]
\begin{center}
\hspace{-0.1cm} \epsfxsize11cm\epsffile{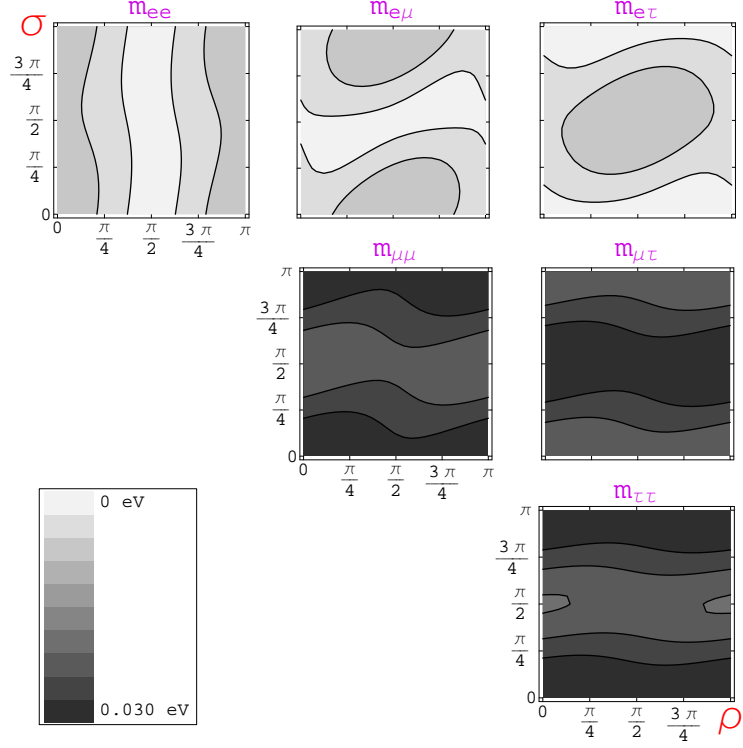}
\caption{The Majorana mass matrix for the
normal mass hierarchy: $m_3/m_2=5$, 
$m_1\approx 0.006$ eV. 
We show contours of  constant mass in 
the $\rho-\sigma$ plots for the moduli of  mass matrix elements.  
We take  for other parameters $\Delta m^2_{12}=7 \times 10^{-5}
{\rm eV}^2$, $\Delta m^2_{13}= 2.5 \times 10^{-3} {\rm eV}^2$, 
$\tan^2\theta_{12}=0.42$, $\tan\theta_{23}=1$, $\sin \theta_{13}=0.1$, and $\delta=0$.}
\label{norm5}
\end{center}
\end{figure}

\begin{figure}[h!]
\begin{center}
\hspace{-0.1cm} \epsfxsize11cm\epsffile{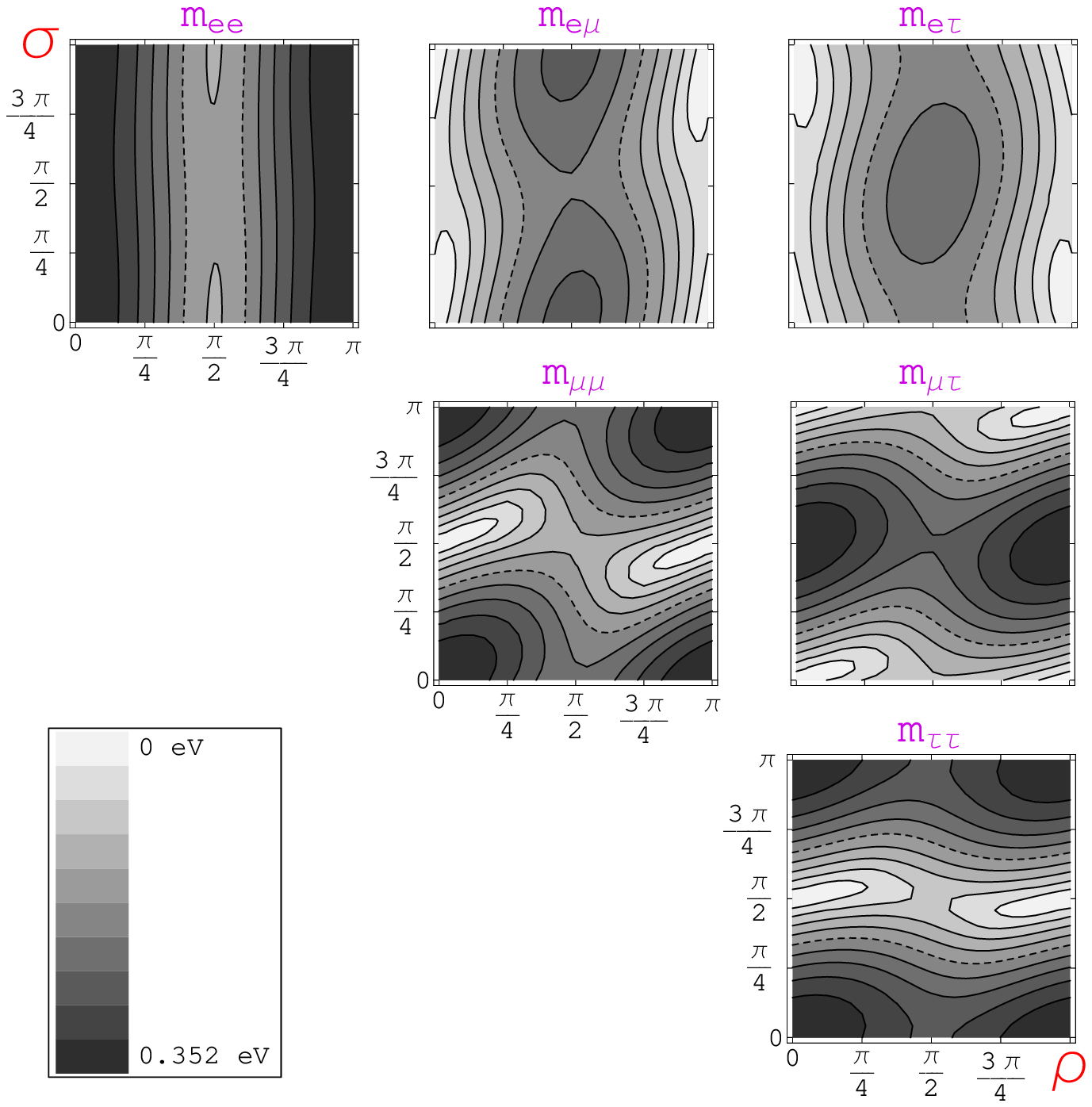}
\caption{The same as in Fig. \ref{norm5} for the quasi-degenerate spectrum:
$m_3/m_2 =  1.01$, $m_1\approx 0.35$ eV.}
\label{deg1}
\end{center}
\end{figure}

\begin{figure}[h!]
\begin{center}
\hspace{-0.1cm} \epsfxsize11cm\epsffile{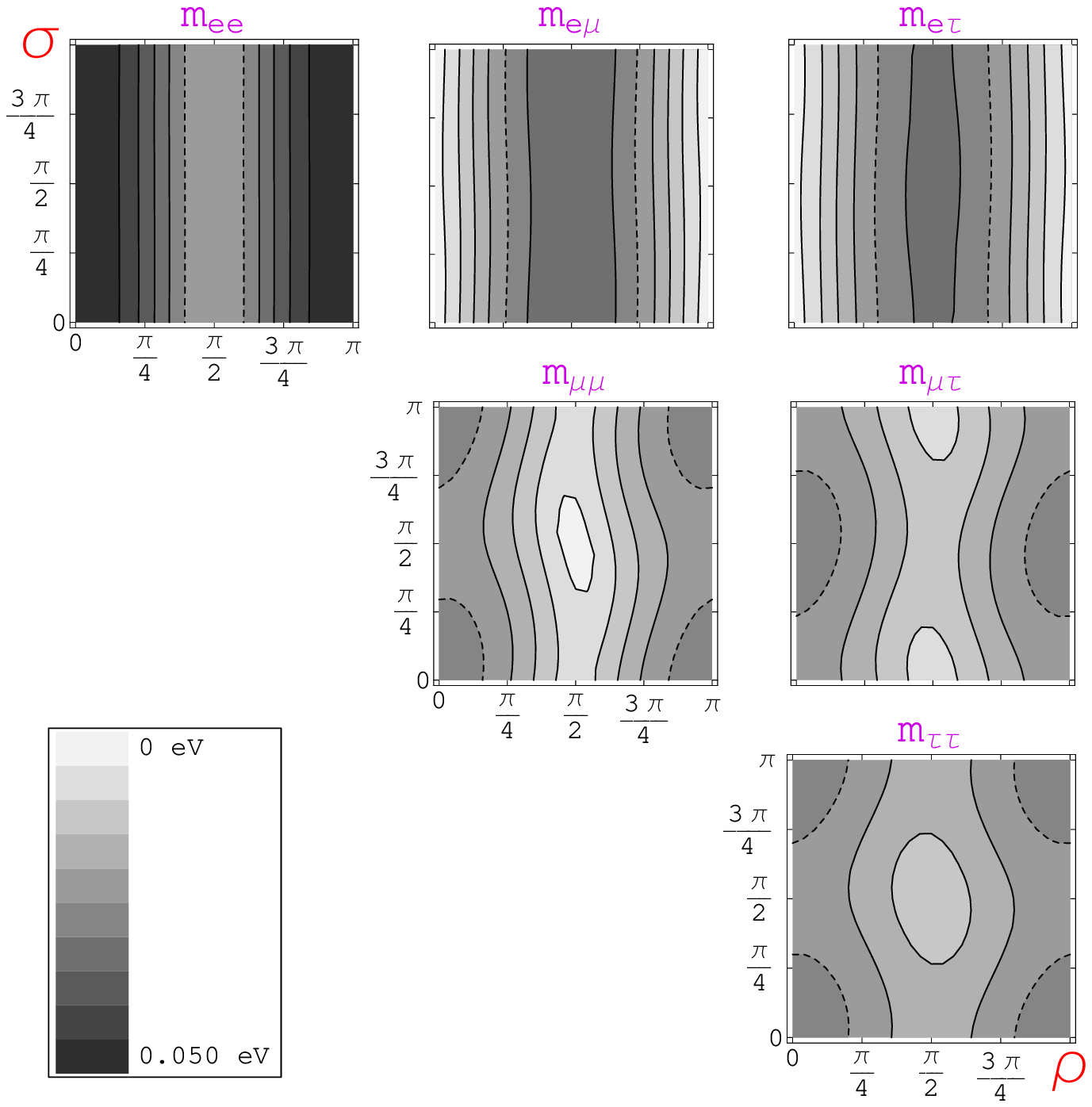}
\caption{The same as in Fig.~\ref{norm5} for the inverted mass hierarchy: 
$m_3/m_2 = 0.1$, $m_3\approx 0.005$ eV.
}
\label{inv01}
\end{center}
\end{figure}

The results of reconstruction of the mass matrix\cite{alta,matrix} are shown in 
Figs:~\ref{norm5},~\ref{deg1}, and~\ref{inv01} 
as the   $\rho - \sigma$ plots for the absolute values of the   
6 independent matrix elements\cite{matrix}.  These figures correspond to  three 
extreme cases: normal mass hierarchy, quasi-degenerate spectrum  
and inverted mass hierarchy. 
The figures illustrate a variety of possible structures. In particular, 
for the normal mass hierarchy (Fig:~\ref{norm5}) there is clear structure with the 
dominant $\mu - \tau$ block.  
Interesting parameterizations of the mass matrix 
(up to an overall mass factor) are 
\begin{equation}
\left(
\begin{tabular}{lll}
0 & 0 & $\lambda$ \\
0 & 1 & 1\\
$\lambda$ & 1 & 1
\end{tabular}
\right),~~~~~ 
\left(
\begin{tabular}{lll}
$\lambda^2$ & $\lambda$ & $\lambda$\\
$\lambda$ & 1 & 1\\
$\lambda$ & 1 & 1\\
\end{tabular}
\right), 
\label{eq:nhier}
\end{equation}
where $\lambda \sim 0.2$. Also the matrix similar to the first one 
in Eq.~(\ref{eq:nhier}) with $m_{12} \sim \lambda$ 
and  $m_{13} \approx 0$ is possible. 
%%%
%%%%%%%%%%%%%%%%%%%%%%%%%%%%%%%%%%ffff9%%%%%%%%%%%%%%%%%%%%%%%%%%%%%%%%%%%%%
%%%%%%%%%%%%%%%%%%%%%%%%%%%%%%%%%%%%%%%%%%%%%%%%%%%%%%%%%%%%%%%%%%%%%%%
%%
%%%%%%%%%%%%%%%%%%%%%%%%%%%%%%%%%%ffff10%%%%%%%%%%%%%%%%%%%%%%%%%%%%%%%%%%%%%
%%%%%%%%%%%%%%%%%%%%%%%%%%%%%%%%%%%%%%%%%%%%%%%%%%%%%%%%%%%%%%%%%%%%%%%

In the case of a quasi-degenerate spectrum,  the interesting dominant structures are 
\begin{equation}
\left(
\begin{tabular}{lll}
1 & 0 & 0\\
0 & 1 & 0\\
0 & 0 & 1\\
\end{tabular}
\right),~~~~~~
\left(
\begin{tabular}{lll}
1 & 0 & 0\\
0 & 0 & 1\\
0 & 1 & 0\\
\end{tabular}
\right). 
\label{eq:degen}
\end{equation}
These matrices are realized for values of  phases in the corners of the plots: 
$\rho, \sigma = 0, \pi$ (the first matrix) or at $\rho = 0, \pi$, 
$\sigma = \pi/2$ (the second one) which  corresponds to definite CP-parities of 
the mass eigenstates. 
Also the ``democratic" structure with equal 
moduli of elements is possible for the non-trivial 
values of phases\rlap{.}\,\cite{demo} 
Changing the phases one can get any 
intermediate structure between those in Eqs.~(\ref{eq:nhier}) and (\ref{eq:degen}). 
%%%%%%%%%%%%%%%%%%%%%%%%%%%%%%%%%%ffff11%%%%%%%%%%%%%%%%%%%%%%%%%%%%%%%%%%%%%
%%%%%%%%%%%%%%%%%%%%%%%%%%%%%%%%%%%%%%%%%%%%%%%%%%%%%%%%%%%%%%%%%%%%%%%

In the case of the inverted hierarchy the $ee$- element is not small generically. 
Among  interesting (and very different) examples are
\begin{equation}
\left(
\begin{tabular}{lll}
0.7 & 1 & 1\\
1 & 0.1 & 0.1\\
1 & 0.1 & 0.1\\
\end{tabular}
\right), ~~~~~~~
\left(
\begin{tabular}{lll}
1 & $0.1$ & $0.1$\\
$0.1$ & 0.5 & 0.5\\
$0.1$ & 0.5 & 0.5 \\
\end{tabular}
\right). 
\label{eq:ihier}
\end{equation}

Notice that there are clear advantages to consider the mass matrix instead of 
oscillation parameters: 

1).  The mass matrix unifies information contained in masses 
and mixing angles and this may provide some deeper insight to the underlying 
theory. 

2). The elements  of the mass matrix are physical parameters: 
they can be immediately measured in the 
neutrinoless beta decay and, in principle,  in other similar processes with   
$\Delta L = 2$. 

3). In the SM and MSSM the radiative corrections 
(the renormalization group effects) on the mass matrix 
are very small, so  they do not change the 
structure of the mass matrix up to  
the scale where the corresponding mass operators are formed or up to the  symmetry scale.  

The disadvantage is that 
the flavor basis may differ from the symmetry basis,  
where the symmetry (as well as mechanism of symmetry violation) 
and underlying dynamics are realized.

%%%%%%%%%%%%%%%%%%%%%%%%%%%%%%%%%%%%%
\subsection{Observations}
%%%%%%%%%%%%%%%%%%%%%%%%%%%%%%%%%%%%%%%%%

Scanning the $\rho-\sigma$ plots shown in 
Figs.~\ref{norm5},~\ref{deg1}, and~\ref{inv01},  one can make the 
following observations. 

1). A large variety of different structures is still possible, depending strongly on 
the unknown $m_1$, type of mass hierarchy and  Majorana phases.  

2). Generically the hierarchy of elements is not strong: within 1 order of magnitude. 
At the same time,  matrices with one or two exact 
zeros are not excluded\rlap{.}\,\cite{zero} 

3). Typically,  the hierarchical structures appear 
for the Majorana phases near 0, $\pi/2$, or  $\pi$.

4). Matrices  are possible with: 

- dominant (i) diagonal elements $(\sim I)$, (ii)  $\mu \tau$-block, 
(iii) $e$-row elements, (iv) $ee-, \mu\tau-, \tau\mu-$ elements (triangle structure), 

- democratic structure,  

- flavor alignment, 

- non-hierarchical structures with all elements of the same order, 

- flavor disordering, ``anarchy"~\cite{anarc}.

5). Matrices can be parameterized in terms of powers of small parameter $\lambda = 0.2 - 0.3$
consistent with the Cabibbo mixing.

Clearly, at present there is enormous degeneracy 
in structures of the neutrino mass matrices which can reproduce  data. 
The degeneracy will be  be reduced substantially,  if we have more information about  
the absolute mass scale, the mass hierarchy 
and at least one Majorana phase.

%%%%%%%%%%%%%%%%%%%%%%%%%%%%%%%%%%%%%%%%%%%%%%%%%%%%%%%%%%%%%%%%%%%%%
\section{New theoretical puzzle?}
%%%%%%%%%%%%%%%%%%%%%%%%%%%%%%%%%%%%%%%%%%%%%%%%%%%%%%%%%%%%%%%%%%%%%%%%

\subsection{Expected and unexpected}

What is behind the observed structure of the neutrino masses and mixing?
Do we really encounter  new theoretical problem? 
The hope was that neutrinos will reveal something simple which will shed
a light on physics of the high energy scales and on the fermion masses in general.

A plausible scenario was:

\noindent
$\bullet$ seesaw mechanism~\cite{sees};

\noindent
$\bullet$ quark-lepton symmetry,  in a sense that $m_D (neutrino) \sim m(quark)$; 

\noindent
$\bullet$ simple structure of $M_R$ - the mass matrix
of the  RH  neutrinos, {\it e.g.}, $M_R \propto I$ or $M_R \propto m(quarks)$.

This  leads typically  to the hierarchical mass spectrum of light neutrinos and to small
lepton mixing. If this scenario is confirmed, the problem would be probably closed, 
and it would be difficult to add something more.

Instead, unexpected pattern of the lepton mixing  has been  found with
maximal or near maximal 2 - 3 mixing; large   
1 - 2 mixing with, however, significant deviation
from maximal mixing.  Type of the spectrum is not yet clear.
However hierarchy, if exists, is weaker than that for quarks and charged
leptons.

%%
%It is this what gives
%whole excitement  and makes our field alived.
%%

All items of the ``plausible scenario'' are questioned now:

The quark-lepton symmetry? - Less obvious though,  still can be realized at some 
level.

The seesaw mechanism? -  Still is very appealing, though there is no clear
indication from the pattern of mixing.

Simple structure of the mass matrix of the RH neutrinos? - Probably no. 

At this point plenty of various scenarios have been suggested 
with  two extremes which we will discuss next. 

%%%%%%%%%%%%%%%%%%%%%%%%%%%%%%%%%%%%%%%%%%%%%%%%%%%%%%%%%
\subsection{Two extremes: symmetry or no symmetry}
%%%%%%%%%%%%%%%%%%%%%%%%%%%%%%%%%%%%%%%%%%%%%%%%%%%%%%%%%%%%%

\noindent
{\it I. Quasi-degenerate spectrum, maximal 2- 3 mixing}. 
This certainly implies flavor symmetry like $Z_2$,  $A_4$~\cite{A4} or  
$SO(3)$~\cite{so3}.
Lepton (neutrino) mass/mixing pattern 
strongly deviates from quark masses and mixing pattern.
Flavor symmetry of the neutrino sector is broken in the quark and charged lepton 
sectors. 
Oscillation observables $\Delta m_{12}^2$ and $\Delta m_{23}^2$ have
no substantial imprint in the mass matrix: they appear as
small perturbations of the structure determined
by the mixing and the Majorana phases.\\

\noindent
{\it II. Hierarchical spectrum, deviation of the 2-3 mixing from maximal}. 
No special symmetry is needed if 2-3 mixing deviates significantly from
maximal value. In this case 
mixings can be  determined by the condition of ``naturalness" of the
mass matrices according to which the mixing angles satisfy  equalities
\begin{equation}
\tan \theta_{ij} \sim  \sqrt{\frac{m_i}{m_j}}, 
\end{equation}
where $m_i$ are the eigenvalues~\cite{barshay}.

Rotations which follow from diagonalizations of the
up and down mass matrices  cancel each other in the quark sector
thus leading to a small quark mixing  and they sum up
in the leptonic sector~\cite{barshay}. The later can be related somehow to
the mechanism which leads to a smallness of the neutrino masses.
In particular, for the 2-3 mixing we have
\begin{equation}
\theta_{23} \sim \sqrt{\frac{m_2}{m_3}} + \sqrt{\frac{m_{\mu}}{m_{\tau}}}
\sim 38^0 
\label{}
\end{equation}
which  is well within the allowed region. 
In this case the oscillation parameters are well imprinted into
the structure of mass matrix. The Majorana phases are not important.\\
Notice that the present atmospheric neutrino data indeed give some hint of deviation 
of the 2-3 mixing from maximal as it was discussed in Sec. 2.2.

%%%%%%%%%%%%%%%%%%%%%%%%%%%%%%%%%%%%%%%%%%%%%%%%%%%%%%%%%%%%%
\subsection{Large Mixing and Degeneracy}
%%%%%%%%%%%%%%%%%%%%%%%%%%%%%%%%%%%%%%%%%%%%%%%%%%%%%%%%%%%%%

Is large mixing implies  degeneracy of the neutrino spectrum?
Or {\it vice versa} does degeneracy of the neutrino mass spectrum
explains large or maximal mixing? 

Let us consider the two  generation case and  introduce
the degeneracy parameter 
\begin{equation}
\delta_{23}  \equiv \frac{\Delta m}{m} \approx \frac{\Delta m^2}{2m^2}, 
\end{equation}
as well as  the parameter which characterizes the deviation of 2-3 mixing from
maximal (\ref{devia}) 
%\begin{equation}
%D_{23} \equiv 1/2 - \sin^2\theta_{23}.
%\end{equation}
Only in one case a  strong degeneracy is related to
small deviation from maximal mixing: this happens for the pseudo-Dirac neutrinos
with mass matrix
\begin{equation}
\left(
\begin{tabular}{ll}
0 & 1 \\
1 & $\epsilon$\\
\end{tabular}
\right), 
\label{eq:ihier}
\end{equation}
where $\epsilon \ll 1$. In this case $D_{23} \approx \delta_{23}$. 
For $m \sim 0.25$ eV, the atmospheric splitting implies
$\delta_{23} \sim 0.03$. Therefore if $D_{23} > 0.01$ is established,
the possibility of Eq.~(\ref{eq:ihier}) will be excluded.

In  other cases the deviation and the mass split are not
related. For instance, the matrix
\begin{equation}
\left(
\begin{tabular}{ll}
1 & $\epsilon$ \\
$\epsilon$ & 1\\
\end{tabular}
\right)
\label{eq:ihier1}
\end{equation}
gives exactly maximal mixing $D_{23} = 0$ but arbitrary mass split:
$\delta_{23} = 2 \epsilon$.
In contrast, the matrix
\begin{equation}
\left(
\begin{tabular}{ll}
$\epsilon$ & 1 \\
1 & - $\epsilon$\\
\end{tabular}
\right)
\label{eq:ihier2}
\end{equation}
corresponds to zero mass split, $\delta_{23} = 0$, but arbitrary mixing
$D_{23} = \epsilon/2$.

Simple relations between degeneracy and deviations is absent
also due to the presence of

\begin{itemize}

\item
third neutrino, 

\item
Majorana phases.

\end{itemize}

Furthermore, in the case of the degenerate spectrum for the first and second 
generations, smaller
splitting, $\delta_{12} \sim 10^{-3}$,  is associated to larger deviation 
$D_{12} \sim 0.2$.

%%%%%%%%%%%%%%%%%%%%%%%%%%%%%%%%%%%%%%%%%%%%%%%%%%%%%%
\subsection{No simple structure?}
%%%%%%%%%%%%%%%%%%%%%%%%%%%%%%%%%%%%%%%%%%%%%%%%%%%%%%%%%%

Our attempts to find simple structures for the neutrino masses matrices
may fail for the following reasons.

1). Neutrino mass matrix can obtain several  relevant contributions from
new physics at all possible scales, $M_a$, from the EW scale to the Planck
scale. As a consequence, the structure of the mass matrix can be rather complicated.
The effective operator at low energies 
(after integrating out the heavy degrees of freedom) 
can be written as\,\cite{eff}
\begin{equation}
\sum_a \frac{\lambda_{ij}^{a}}{M_a} (L_i H)^T(L_j H), ~~~ i,j = e, \mu, \tau ,
\label{effop}
\end{equation}
where $L_i$ is the lepton doublet,  
$\lambda_{ij}^{a}$  are the dimensionless couplings. 
After the EW symmetry breaking it generates the  neutrino masses
\begin{equation}
m_{ij} =  \sum_a \frac{\lambda_{ij}^a \langle H \rangle^2}{M_a}. 
\label{masssum}
\end{equation}
The sum is crucial here. 
For $\lambda_{ij}^{a} \sim 1$ and $ M = M_{Pl}$ we find $m_{ij} \sim 10^{-5}$~eV\rlap{.}\,\cite{planck}
Even this contribution can be relevant for the sub-leading
structures of the mass matrix and phenomenology.

2). The presence of new (sterile) neutrino states can have dramatic consequences.
Heavy sterile neutrinos ($M > 1$ keV) may contribute to the sum
(\ref{masssum}) modifying substantially mixing of the active neutrinos and mass
splitting~\cite{bal}.  Light sterile neutrinos  ($m < 1 eV$ ) can change 
low energy phenomenology and therefore determination of the neutrino
(oscillation)  parameters.
Therefore to understand the neutrino masses and mixing we need to restrict or
control possible effects of sterile neutrinos.

%%%%%%%%%%%%%%%%%%%%%%%%%%%%%%%%%%%%%%%%%%%%%%%%%%%%%%%%%%
\section{Conclusions}
%%%%%%%%%%%%%%%%%%%%%%%%%%%%%%%%%%%%%%%%%%%%%%%%%%%%%%%%%%%%%%%%%%%%%%%

\noindent
1. Enormous progress has been achieved in
the determination of the neutrino masses and mixings,  
in reconstruction of the neutrino mass spectrum,  and in studies of the neutrino mass matrix.\\

\noindent
2. Achieved results allow us to formulate clear program of further experimental
and phenomenological studies.\\

\noindent
3. Amazing pattern of the lepton mixing emerges which probably composes
the new theoretical puzzle. Its resolution  may lead us to new fundamental results,
to discovery of new symmetries of Nature.\\

\noindent
4. From this theoretical perspective, the most important future measurements turn 
out to be 

\begin{itemize}

\item

determination of the absolute mass scale, and tests of degenerate
scenario;

\item
searches for deviation of the 2-3 mixing from maximal value; 

\item

measurements of 1-3 mixing,  and of course,  

\item
establishing the Majorana nature of neutrinos. 

\end{itemize}

\section{Acknowledgments}

I am grateful to E. Kh. Akhmedov,  P. de Holanda, M. Frigerio, C. Lunardini  
and O. Peres  for fruitful discussions 
and contributions to this talk.  Many thanks to Milla for this marvelous meeting. 

%\section{First Appendix}

\renewcommand{\theequation}{A.\arabic{equation}}

%%%%%%%%%%%%%%%%%%%%%%%%%%%%%%%%%%%%%%%%%%%%%%%%%%%%%%%%%%%%%%%%%%%%%%%%%%%%%%%%%%%%%%%%%%%%%%%%

%\balance

\end{document}